\documentclass[sigconf,screen,authorversion]{acmart}

\usepackage{subcaption}
\usepackage[skip=3pt]{caption}
\usepackage{graphicx}
\usepackage{amsthm}
\usepackage[inline]{enumitem}
\usepackage{multirow}
\usepackage[capitalize,noabbrev]{cleveref}

\theoremstyle{definition}

\AtBeginDocument{%
  \providecommand\BibTeX{{%
    \normalfont B\kern-0.5em{\scshape i\kern-0.25em b}\kern-0.8em\TeX}}}

\makeatletter
\newcommand\headingnodot{\def\@toclevel{4}%
  \@startsection{paragraph}{4}{\z@}%
  {-.2\baselineskip \@plus -2\p@ \@minus -.2\p@}%
  {-3.5\p@}%
  {\ACM@NRadjust{\bfseries}}}
\newcommand{\heading}[1]{\headingnodot{#1.}}
\makeatother

\begin{document}

\author{Santiago de Leon-Martinez}
\affiliation{%
  \institution{Brno University of Technology}
  \city{Brno}
  \country{Czechia}
}
\additionalaffiliation{%
  \institution{Kempelen Institute of Intelligent Technologies}
  \city{Bratislava}
  \country{Slovakia}
}
\email{santiago.deleon@kinit.sk}
\orcid{0000-0002-2109-9420}

\author{Robert Moro}
\affiliation{%
  \institution{Kempelen Institute of Intelligent Technologies}
  \city{Bratislava}
  \country{Slovakia}
}
\email{robert.moro@kinit.sk}
\orcid{0000-0002-3052-8290}

\author{Branislav Kveton}
\orcid{0000-0002-3965-1367}
\affiliation{%
  \institution{Adobe Research}
  \city{San Jose, CA}
  \country{United States}
}
\email{kveton@adobe.com}

\author{Maria Bielikova}
\orcid{0000-0003-4105-3494}
\affiliation{%
  \institution{Kempelen Institute of Intelligent Technologies}
  \city{Bratislava}
  \country{Slovakia}
}
\email{maria.bielikova@kinit.sk}

\renewcommand{\shortauthors}{de Leon-Martinez et al.}
\settopmatter{authorsperrow=2,printfolios=true}

\copyrightyear{2026}
\acmYear{2026}
\setcopyright{cc}
\setcctype{by}
\acmConference[IUI '26]{31st International Conference on Intelligent User Interfaces}{March 23--26, 2026}{Paphos, Cyprus}
\acmBooktitle{31st International Conference on Intelligent User Interfaces (IUI '26), March 23--26, 2026, Paphos, Cyprus}
\acmDOI{10.1145/3742413.3789166}
\acmISBN{979-8-4007-1984-4/2026/03}


\title[Riding the Carousel]{Riding the Carousel: The First Extensive Eye Tracking Analysis of Browsing Behavior in Carousel Recommenders}

\begin{abstract}
Carousels have become the de-facto standard user interface in online services. However, there is a lack of research in carousels, particularly examining how recommender systems may be designed differently than the traditional single-list interfaces. One of the key elements for understanding how to design a system for a particular interface is understanding how users browse. For carousels, users may browse in a number of different ways due to the added complexity of multiple topic defined-lists and swiping to see more items.

Eye tracking is the key to understanding user behavior by providing valuable, direct information on how users see and navigate. In this work, we provide the first extensive analysis of the eye tracking behavior in carousel recommenders under the free-browsing setting. To understand how users browse and model their behavior, we examine the following research questions : 1) where do users start browsing, 2) how do users transition from item to item within the same carousel and across carousels, and 3) how does genre preference impact transitions? 

This work addresses a gap in the field and provides the first extensive empirical results of eye tracked browsing behavior in carousels for improving recommenders. Taking into account the insights learned from the above questions, our final contribution is to provide takeaways for carousel recommender system designers to better optimize their systems for user browsing behavior. The most important being an improved reordering of the ranked item positions to account for browsing behavior after swiping. These contributions aim not only to help improve current systems, but also to encourage and allow the design of new user models, systems, and metrics that are better suited to the complexity of carousel interfaces.

\end{abstract}


\maketitle

\section{Introduction}
Over the years, recommender systems and information retrieval research have primarily focused on single ranked list interfaces~\cite{hearst-2009-search}. In the late 2000s, research was extended to include different types of interfaces, such as the 2D grid (i.e, single 2D ranked list) that is commonly used for web image search results or movie recommenders~\cite{doi:10.1177/154193120705101831, 10.1145/1743666.1743736}.
Today, \textit{carousel} and \textit{multi-list} interfaces\footnote{Carousel interfaces are a subclass of multi-list interfaces with multiple swipeable lists called carousels. Each carousel has a topic that defines all the items displayed left-to-right with the the possibility to scroll a given carousel horizontally, called \textit{swiping}. Multiple topics and their carousels are displayed vertically in rows.} are encountered by users of e-commerce and streaming services daily, but there is little research focused on these interfaces. While general recommender system and retrieval research can be applied to any interface (e.g. better algorithms, learning better representations, etc.) there is a fundamental question left unanswered for under researched interfaces: \textit{How do users browse and interact with this specific type of interface and based on this how should items be presented?} This is a particularly important question for carousels due to their added complexity allowing for browsing vertically, horizontally, or by topic.

For 1D ranked lists, this question has been answered primarily through eye tracking studies \cite{castagnos_eye-tracking_2010, 10.1145/2959100.2959150, li_towards_2017,granka_eye-tracking_2004, cutrell_what_2007,pan_determinants_2004,hofmann_eye-tracking_2014} that provide direct, rich information of how the user sees/browses the interface. On the other hand, for carousel interfaces, there is only one user study and dataset paper with only preliminary eye tracking results of how users browse \cite{deleon_et_al_2025}. This work, using the same dataset, provides the first extensive analysis of eye tracking results in carousel interfaces. Our goal is to understand how users browse carousel interfaces, model their behavior, and gather insights that can help inform how carousel interfaces can be better designed. 

Our approach to understanding user browsing behavior in carousels is to begin  where users start browsing and then see how they traverse the defining elements of the carousel: carousel rows and item columns. We focus on the following research questions for carousels: 
\begin{enumerate}
    \item[RQ~1] Where do users start browsing?
    \item[RQ~2] How do users transition from carousel row to carousel row?
    \item[RQ~3] How do users transition from item to item (column-to-column) within a given carousel?
    \item[RQ~4] How does genre preference impact carousel row transitions? 
\end{enumerate}
From our analyses and answers of these questions, we then provide suggestions/takeaways to improve the design of carousel recommendation system, namely reordering items to account for swiping behavior and empirical fixation propensity, and encourage research in areas where there is still a lack of understanding of user behavior.

\section{Related Work: User Browsing Behavior}
\label{Related Browsing Behavior}

\heading{Single 1D ranked lists \& search engine results pages}
For single 1D ranked lists and search engine results pages, it was reasonably assumed that users browse the items or results top-down. Top-down browsing behavior was confirmed through both eye tracking studies in recommender systems \cite{castagnos_eye-tracking_2010, 10.1145/2959100.2959150, li_towards_2017} and search results \cite{granka_eye-tracking_2004, cutrell_what_2007,pan_determinants_2004,hofmann_eye-tracking_2014} as well as through empirical click data \cite{joachims_accurately_2016}. Not only did these eye tracking studies allow researchers to test and verify assumptions, but also inspired the cascade click model \cite{craswell_experimental_2008} (based on top-down browsing) and provided learned parameters for click modeling.

\heading{Single 2D ranked lists (slates \& grids)}
Eye tracking studies were also performed to determine browsing behavior in single 2D ranked lists. \citet{10.1145/1743666.1743736} found that rather than browsing line by line (left-right or top-down) users mixed both. This was later called the F-pattern (also known as the \textit{golden triangle}) as users browsed in a shape of the letter F prioritizing the left hand column and going farther right in top rows, while not focusing on the center and especially ignoring items in the bottom right corner \cite{10.1145/2959100.2959150}. However, other studies in grids \cite{doi:10.1177/154193120705101831, doi:10.1177/1541931214581234}, particularly in image search, found other behaviors than the F-pattern, such as middle bias, slower decay, and row skipping \cite{10.1145/3308558.3313514}.

\heading{Multi-list/carousel interfaces}

For item recommendation, \citet{10.1145/3450613.3456809} compared carousels to a single 2D ranked list and found that carousel users were exploring slower and longer and perceived the items as more diverse and novel. In recipe recommendation, \citet{starke_serving_2021} additionally compared 1D single ranked lists to 2D single ranked lists and carousels, finding that carousels and the 2D ranked list  were easier to use than the 1D single ranked lists. \citet{loepp_how_2023} examined user interactions within carousels to find differences in how users perceived and used carousels and that item selection probability was mediated by item position. Similarly, \citet{ferrari_dacrema_offline_2022,felicioni_measuring_2021} proposed NDCG2D as metric for carousel lists based on item positions (vertical, horizontal, and if swiped or not). \citet{rahdari_ranked_2022} proposed the first carousel click model extending the cascade click model called the ``carousel click model,'' in which a users browses topics until finding one of interest and then only examines the items of that topic of interest. Following work was done by carousel user simulation in movie recommendation, finding that carousels were more efficient than single 2D ranked lists \cite{rahdari_towards_2024, 10.1145/3511095.3531278}. 

The first and only eye tracked user study in carousels (as well as the first recommender dataset with eye tracking data included) presented preliminary eye tracking results that users browse unswiped carousels with the F-pattern, but after swiping there appeared to be an reverse browsing pattern (right-to-left rather than left-to-right)~\cite{deleon_et_al_2025}. This work seeks to further analyze the eye tracking data of the aforementioned dataset paper to explain the preliminary results and discover new insights in carousel browsing behavior.

With respect to carousel browsing behavior, there is very little theory available in previous works primarily due to the lack of eye tracking data. As mentioned previously, the only eye tracked results in carousels  \cite{deleon_et_al_2025} show that users browse in an F-pattern and after swiping browse in a reverse F-pattern. All other works rely on small datasets or simulation, without eye tracking data, for testing hypotheses of how users browse. NDCG2D \cite{ferrari_dacrema_offline_2022,felicioni_measuring_2021},  in how it is proposed, implies an F-pattern behavior as the following are considered costly actions: moving further (left-right) in a specific carousel list,  swiping to see the next set of items (which they theorize may be especially costly), and moving top-down from carousel list to carousel list.  \citet{rahdari_ranked_2022} proposed the ``carousel click model'' that models/theorizes that users browse carousel interfaces by looking at the topics first and only examining items of interesting topics. \citet{loepp_how_2023} focused more on user perception, satisfaction, and decision-making measures rather than browsing behavior, and found: 1) that users were more likely to find an item by further exploring carousels rather than seeing more carousels , 2) item position predicts well if it gets selected (supporting NDCG2D), and 3)  they did not find users browsed in the manner described in the ``carousel click model.'' These prior works on carousel browsing behavior, as well as this work, focus on how users interact with the main elements of the carousel, carousel rows and item columns. Our analyses found below also support the F-pattern behavior, help explain the reverse F-pattern, and did not show that users browsed similarly to the ``carousel click model.''

\begin{figure*}[!tb]
    \centering
    \includegraphics[width=0.70\textwidth]{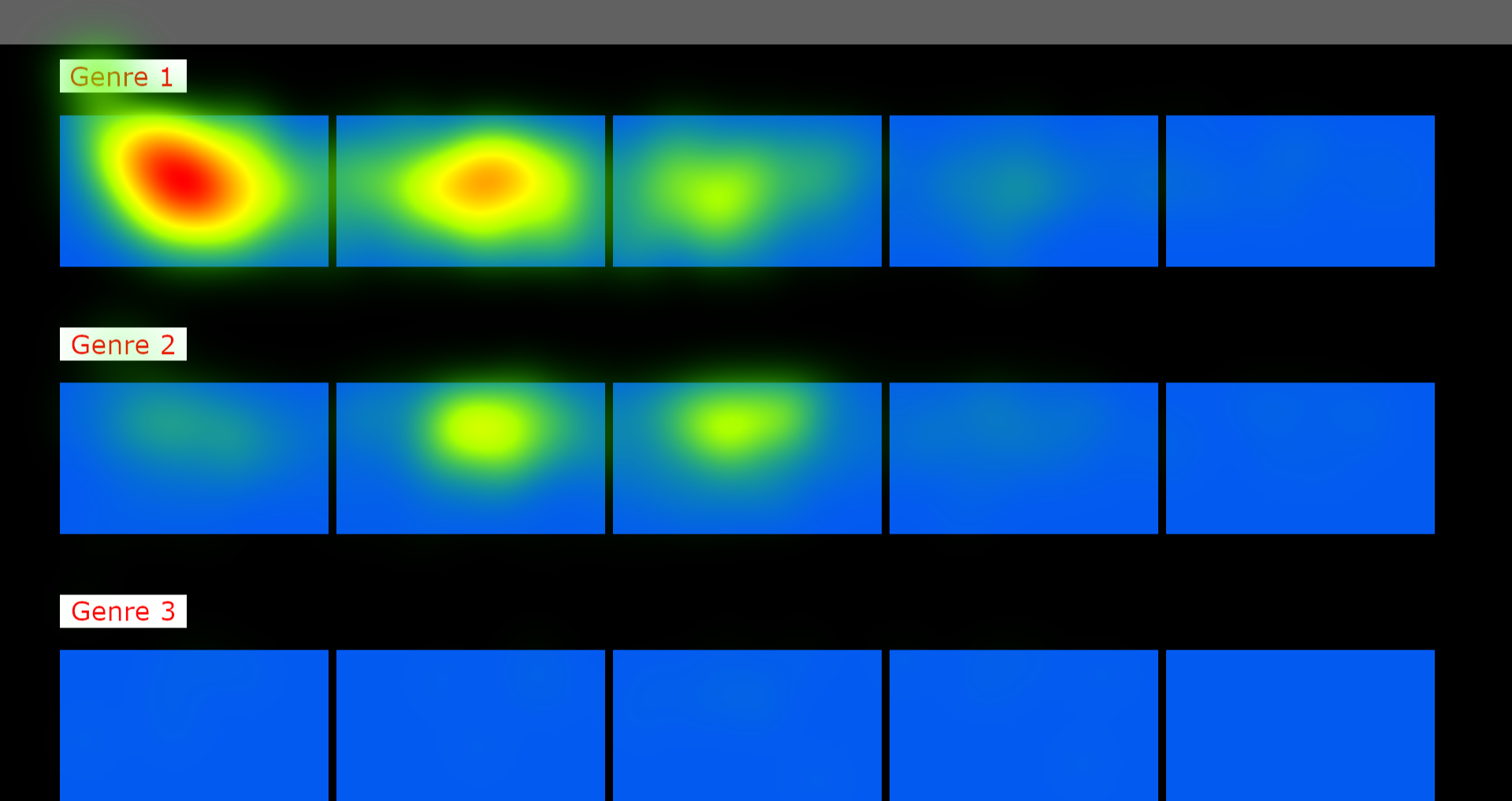} 
    \caption{Aggregate first fixation heat map showing initial browsing bias on a to-scale background (blue boxes represent the 5 visible movie posters in each carousel). The first fixations using x,y pixel position from all users and free-browsing screens are aggregated from more to less fixations (red > orange > yellow > green).}
    
    \label{fig:Initial_bias}
\end{figure*}
\label{results}

\section{Methodology: User Study and Dataset}
\heading{User Study} The available eye tracking and user interaction (clicks, cursor movements, and selection explanations) dataset\footnote{Accessible at: \url{https://zenodo.org/records/15270518}} used for this analysis is taken from an eye tracked user study in carousel movie recommendation~\cite{deleon_et_al_2025} for desktop. Users were eye tracked over 40 different movie selection screen: 30 of free-browsing (pick any movie on the screen),  5 of semi-free browsing (pick a movie from your favorite genre), and 5 of direct search (find a specific movie). This analysis only uses data from the 30 free-browsing screens, as this best emulates the common use case of carousels on a homepage.

The design of the interface was inspired by the Netflix homepage allowing familiar users to browse comfortably and rely on previously learned browsing habits/behaviors. Ten genres (Action, Animation, Comedy, Crime, Drama, Fantasy, Horror, Romance, Sci-Fi, and Thriller) were used to create 10 genre carousels (we later refer to the top-most carousel on the screen as Row 1 and so on) populated with movies taken from the IMDB non-commercial dataset sorted by popularity (number of votes). Each of the genre carousels contained a total of 15 movies, 5 of which are initially present when the carousel is in view and the other two sets of 5 can be arrived to with right or left swiping. All 15 movies (later referred to as Columns 1 through 15) were sorted by popularity meaning that in the initial (unswiped) set of movies the most popular (Column 1) is the first, left-most movie and popularity decreases until the last, right-most movie (Column 5). After right swiping, 5 new movies replace the previous and the new first, left-most movie (Column 6) is the next popular after the last movie of the initial set and popularity decreases as before from left to right. 

A to-scale mock-up of what the interface looks like at initial presentation (only the top 3 genre carousels are initially visible, while the other 7 must be vertically scrolled to) is shown in \cref{fig:Initial_bias} and a sample gif screen recording is available in the dataset paper GitHub\footnote{\url{https://github.com/santideleon/RecGaze_Dataset}}. Across the 30 screens, all genres are randomly ordered while also guaranteeing that:
\begin{enumerate*}[label=(\roman*),leftmargin=*]
    \item each genre carousel is equally present within three initial presentation locations; and 
    \item each genre carousels is equally present in each of the 10 possible row positions on the screens. 
\end{enumerate*}

\heading{Dataset} The free-browsing dataset includes 87 users and 30 movie selection screens totaling 2,610 screens, of which 52 (2\%) screens were omitted from the analysis due to missing data. Of the total 87 users,  47 were men and 39 woman with the following ages: (18--19): 1, (20--29): 50, (30--39): 29, (40--49): 6, and (60--69): 1. 61 participants were gathered in Bratislava while the remaining 26 were in Amsterdam. Genre preference information (along with other user information) was gathered through surveys in 3 different formats: marking preferred / non-preferred for each, 1-star to 5-star rating for each, %
and pick the top or most preferred genre. User perception and satisfaction measures were also gathered (see \cite{deleon_et_al_2025}) along with a psychological decision-making scale, but we do not include these in our analyses as we focus on addressing the open question of how users browse carousel interfaces.

The gaze (raw eye tracking) data was gathered by Tobii 4C remote bar eye tracker (90~Hz sampling rate and error of less than 1 degree~\cite{taore_limits_2024,gibaldi_evaluation_2017}) and pre-processed into fixation data (see \cite{deleon_et_al_2025} for more detail). \textit{Fixations} are stable, fixed points during the gaze or eye movements that are associated with cognitive processes of examining an item. In other words, the eye tracker provides timestamped 90Hz gaze data with x,y pixel screen coordinates representing where the user is looking on the interface, but gaze data is very high frequency and not necessarily representative of cognitive processes (i.e., it is too high frequency to know if the user is truly processing and observing the item at that point in screen). The gaze is processed into fixations (i.e. classified into segments where the user is gazing at a small area within a time window; fast and long distance eye movements are ignored) that represent a user examining one area/item with x,y pixel screen coordinates and a duration. Using the x,y pixel screen coordinates of a fixation, we can find the item/scene of the interface the user is examining.

The dataset is already pre-processed with all fixations labeled with their associated area of interest (AOI), or the item/scene the user is looking at at that moment. The relevant AOIs used for labeling are: movie poster, genre topic text, left/right swipe button, and background. These AOIs have x,y pixel screen bounding boxes based on the interface and the user scroll position. If a fixation x,y screen position is inside of an AOI bounding box then the fixation is associated with that AOI. We focus all of our analyses on these fixated AOIs, including transitions between them, and first fixation x,y positions, which are indicative of how a user browses the whole carousel page from start to finish. Other gaze metrics (such as fixation duration or dwell time) may be more indicative of users’ interest in (attraction of) particular movie items, which is beyond the scope of the current work.

\section{Results: How users browse the carousel}
To understand how users browse for items in carousel recommenders, we begin by examining how users start browsing and the initial positional biases. Then we examine  carousel-to-carousel (row) transitions, item-to-item (column) transitions, and a joint analysis of both transitions only considering the nearest item (i.e. no skipping). Finally, we examine how genre preference may impact this behavior through 3 different methods (preferred vs non-preferred, top genre vs rest, and 5-point rating).

\subsection{RQ 1: Where do users start browsing?}

The aggregate fixation heat map of the first fixation for all users and all free-browsing screens is shown in \cref{fig:Initial_bias}.\footnote{The analysis in \cref{fig:Initial_bias} uses the actual x,y pixel position showing where users were looking, similar to the aggregate all fixation heat map in \cite{deleon_et_al_2025}. All following figures and analyses do not use pixel position, but instead fixations are classified by AOI.} The first fixation is the first available fixation in the dataset or the first part of the interface the user examines. Pre-processing was performed to remove all data before the page finished loading. %
We additionally examined the second and third fixations and a joint first three fixation aggregate heat maps and found no differences, so we only show the first fixation heat map. These 3 heat maps being the same is a reasonable result, as users are likely to examine and fixate on the same item or area as they explore the first item; even if that were not the case then they would likely transition to a nearby item in the same top left corner (indistinguishable in aggregate heat maps).

The heat values in \cref{fig:Initial_bias} show that users are likely to begin with the top left item, which follows the F-pattern behavior. Of note, there is only some focus on the Genre 1 label meaning that the poster images are more likely to draw in the gaze of the user rather than the genre topic. Moreover, it is also possible to see the genre label in the peripheral vision or during scanning and is less likely to be fixated on in general.

User are not always starting with the top left item, which is to be expected when there are visual differences and biases present in the differing movie posters. Despite these visual biases, the positional bias of this top left quadrant is prominent. There are very little to none fixations on the 4th or 5th item of the first two rows and all items of the third row. In summary, these results are not surprising as they are in line with the F-pattern browsing behavior.

\begin{figure*}[!tb]
    \begin{subfigure}{0.7\textwidth}
        \includegraphics[width=.99\linewidth]{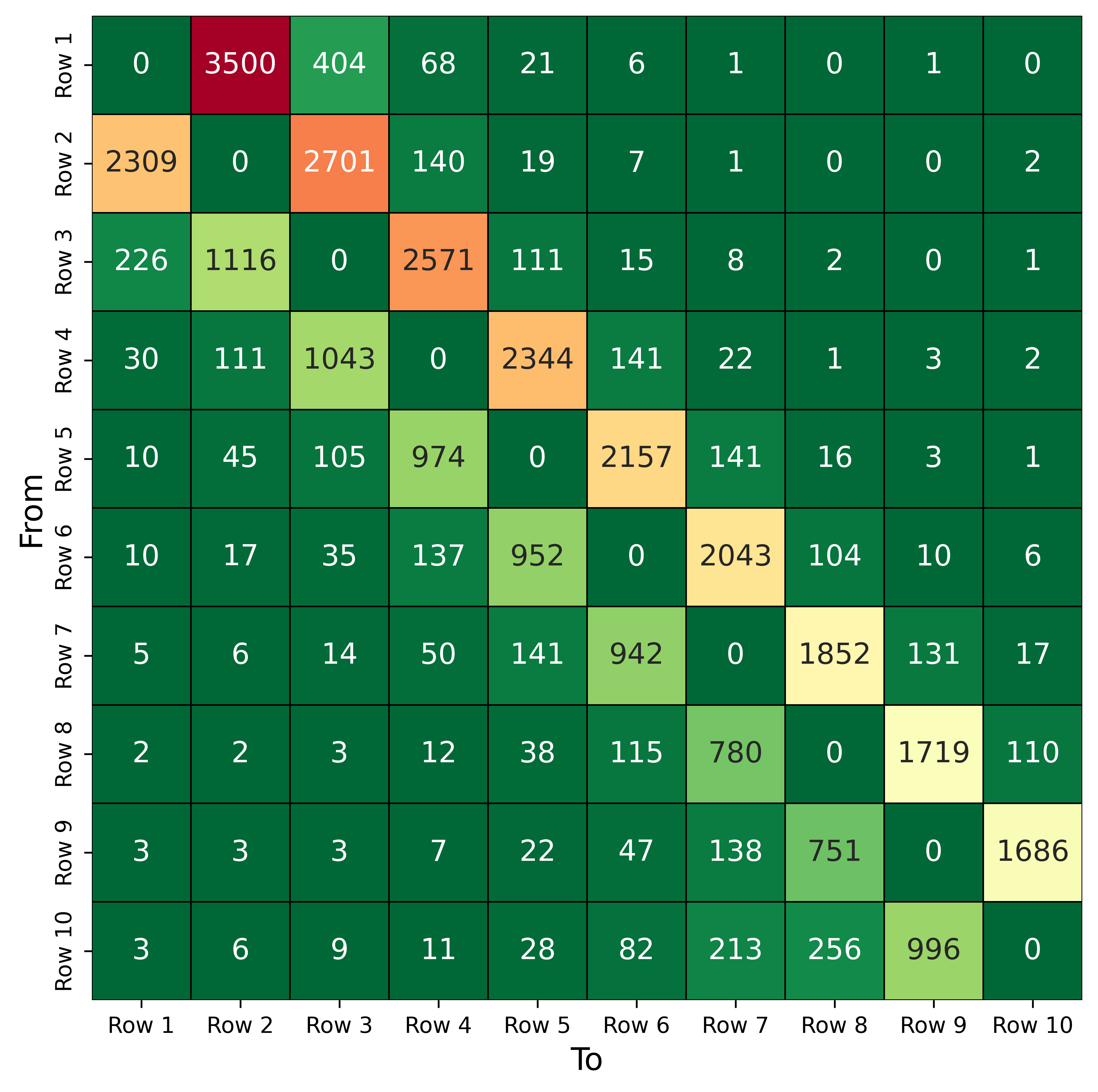}

    \end{subfigure}
    \begin{subfigure}{0.25\textwidth}
    \centering
        \includegraphics[clip, trim =65mm 0mm 0mm 0mm, height = 3cm]{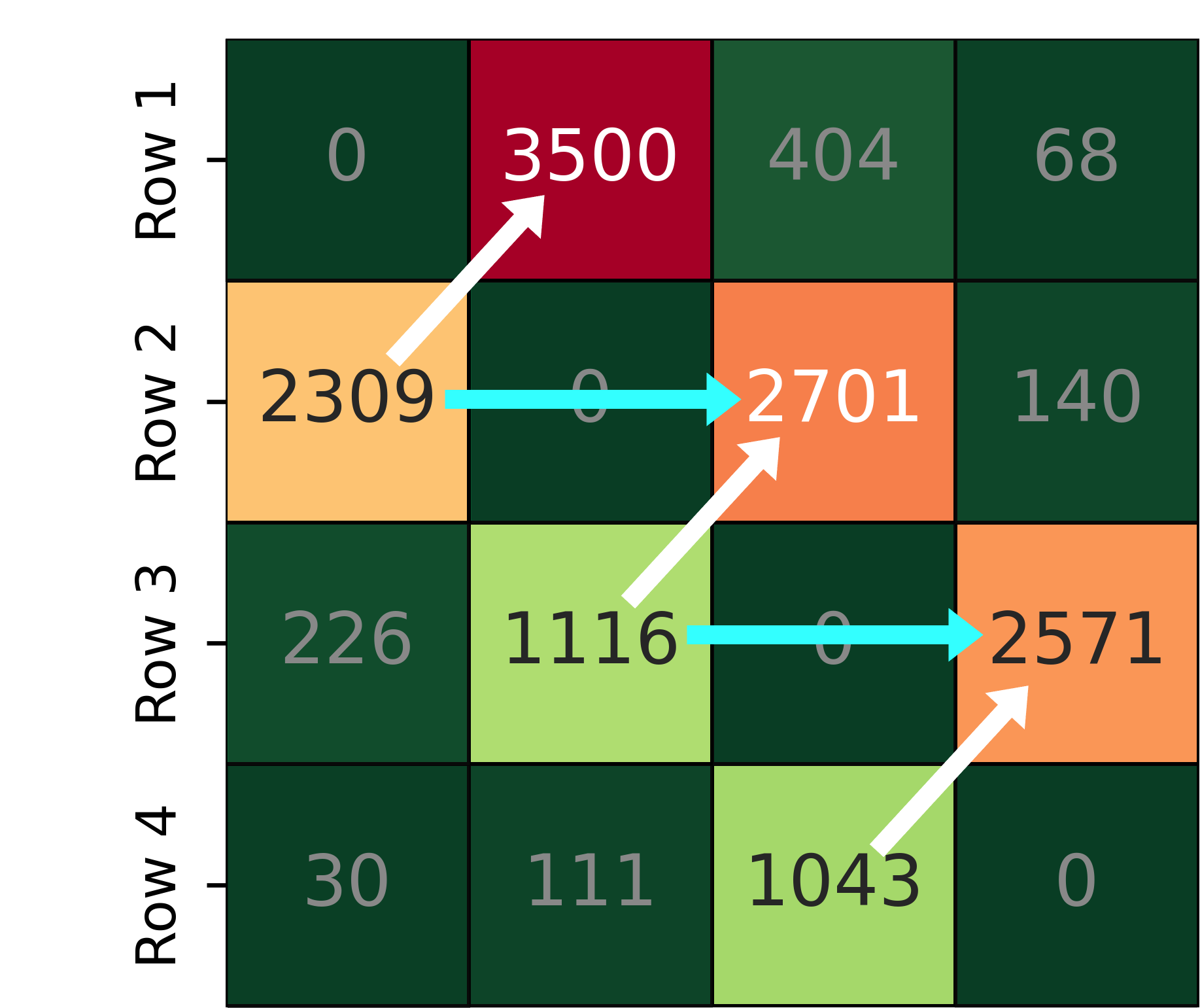}
        \caption{
        Users browse top-down more than bottom-up. The upper diagonal is top-down browsing while the lower diagonal is bottom-up.
        }
        \includegraphics[clip, trim =65mm 0mm 0mm 0mm, height = 3cm]{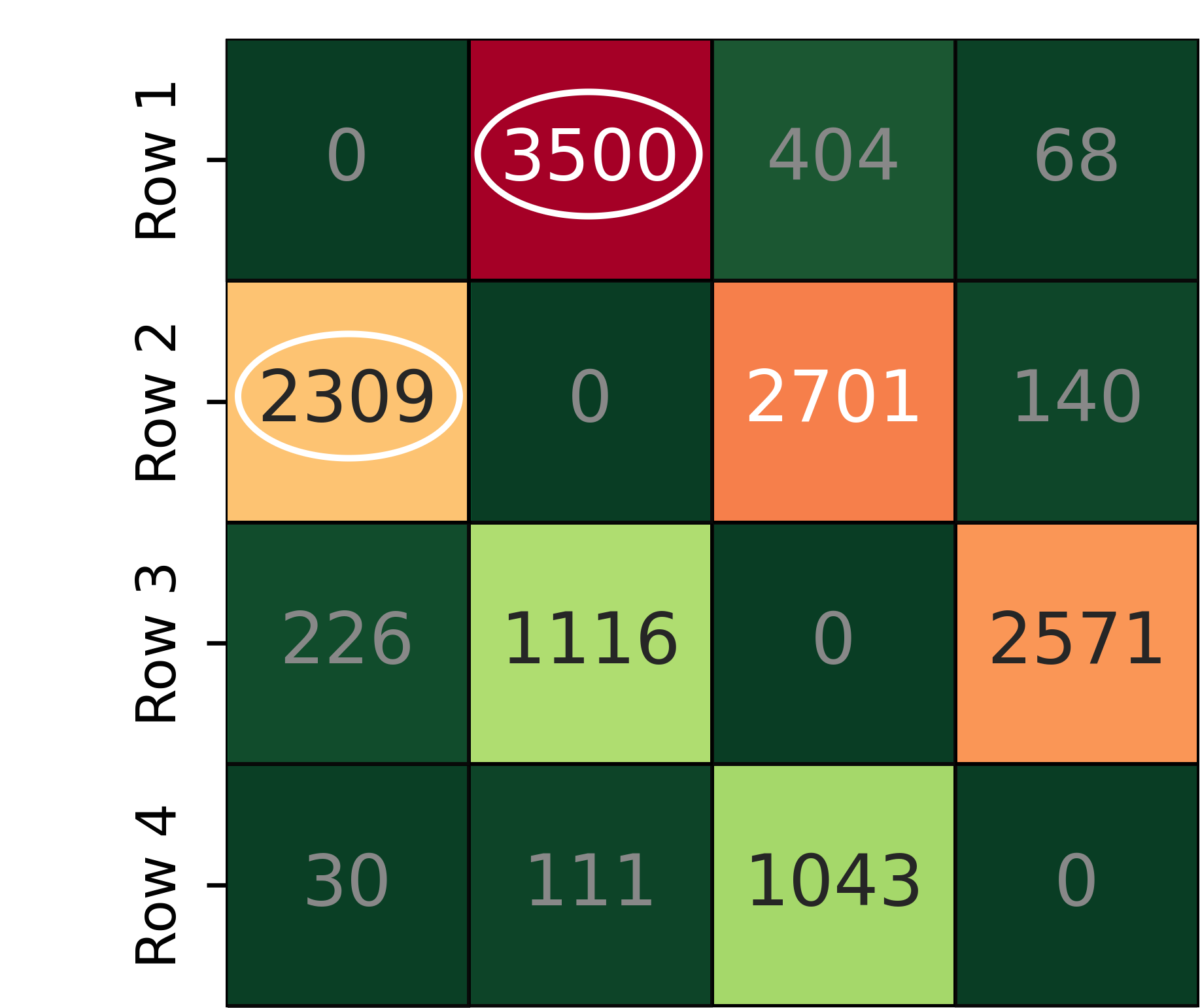}
        
        \caption{Top 2 row bias. %
        }
        \includegraphics[clip, trim =65mm 0mm 0mm 0mm, height = 3cm]{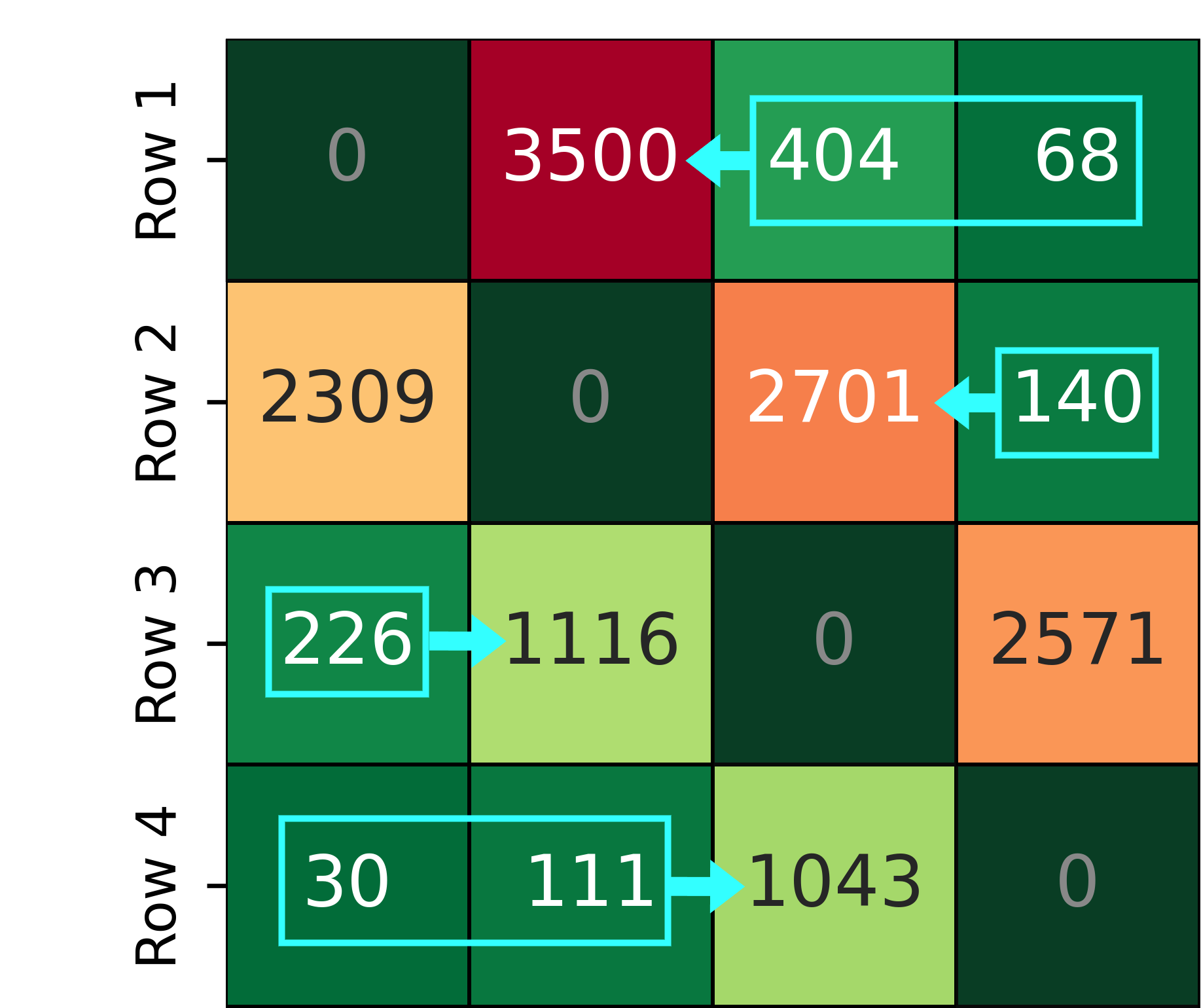}
        
        \caption{Row skipping is less frequent than adjacent transitions. %
        \vspace{0.25cm}
        }

    \end{subfigure}
    
    \caption{Row fixation transitions matrix for all users and free-browsing screens. It shows the number of instances users examined one row (a carousel genre/list) and afterwards examined a different row. Fixations on a genre label and the swipe button (of that row) as well as the items (swiped/unswiped) in that carousel are considered part of the same row. All fixations on the background were ignored (e.g. Row 1 to background to Row 2 is counted as Row 1 to Row 2). (a), (b), (c) visualize the main results in subareas of the matrix. Blue arrows compare transitions from a given row (to a previous/above row vs. to a subsequent/below row) and white arrows compare inverse transitions (from a given row to the previous one vs. from the previous one to the given row). The arrow points towards a higher value, i.e. it indicates a more prevalent pattern / behavior.}

    \label{fig:row_transitions}
\end{figure*}

\subsection{RQ 2: How do users transition across carousels (row-to-row)? }

In \cref{fig:row_transitions}, we show the number of fixation transitions from one carousel genre row to another carousel genre row. A fixation transition occurs when a user fixates or examines an area of the interface and then examines a different area of the interface. Thus, the matrix is computed by counting the number of times a user was fixating or inspecting an AOI of a carousel row (movie, swipe button, or genre label) and then transitioned to examine another AOI of a different carousel row (i.e. all AOIs of a given carousel are the same and we count changes to any AOI of a different carousel row). In other words, this shows how users browse vertically and jump from genre carousel row to genre carousel row including if they are skipping. Note that on 5.9\% of screens, users selected movies in the initial row that they observed without moving to other rows and thus their transitions counts would be zero.

First we compare adjacent transitions, of which there are two types: 1) transitions from a given row (blue arrows in \cref{fig:row_transitions}a), for example given Row 3 we compare the transitions Row 3 to 2 (Row 3-2) and Row 3 to 4 (Row 3-4), and inverse transitions (white arrows in \cref{fig:row_transitions}a), such as comparing Row 2-3 and Row 3-2.  \cref{fig:row_transitions}a shows the result of top-down browsing being more frequent than bottom up browsing through arrows that point to the top-down browsing transitions (upper diagonal) that are always greater than bottom-up transitions (lower diagonal). Both row transitions (blue arrows) and inverse transitions (white arrows) show this pattern throughout the whole diagonal of the matrix.
These results are in line with the vertical top-down browsing behavior that is present in the F-pattern.

For a given row, users are about twice as likely to go down than up except for the case of Row 2 (e.g. Row 2-1: 2309 vs. Row 2-3: 2701). Moreover, visualized in \cref{fig:row_transitions}b the down transition Row 1-2 and up transition 1-2 are inflated when compared to similar down and up transitions. We hypothesize that this is due to a top 2 row bias, which we confirm in a later analysis (see \cref{fig:heat_map}a).

While adjacent transitions are much more common than row skipping transitions (see \cref{fig:row_transitions}c), row skipping is still present in the data. For Rows 3 through 9, the ratio of the count of skip transitions to adjacent transitions is around 1:10. The last Row 10 has the most relative skips around 1:2, but it is also in this case where a user is most likely to skip up to a previous row that is the most interesting (assuming that they sequentially observed all or some of the rows before 10). The other edge row, Row 1 has a ratio of 1:7, which is more likely than Rows 3 to 9. Finally, Row 2 is an outlier with a ratio of 1:33 due to the large number of adjacent transitions (top 2 row bias) and the initial presentation of the page (see \cref{fig:Initial_bias}) that requires scrolling before it is even possible to perform a row skip.

In summary, the row transitions reveal that: 
\begin{enumerate*}[label=(\roman*)]
\item in general, users browse rows top-down, but
\item there is top 2 row bias with users being likely to transition in between the top 2, and
\item row skipping is rare when compared to non-skipping except for the last row. 
\end{enumerate*}

\begin{figure*}[!tb]
\centering
    \begin{subfigure}{0.7\textwidth}
        \includegraphics[width=1.0\linewidth]{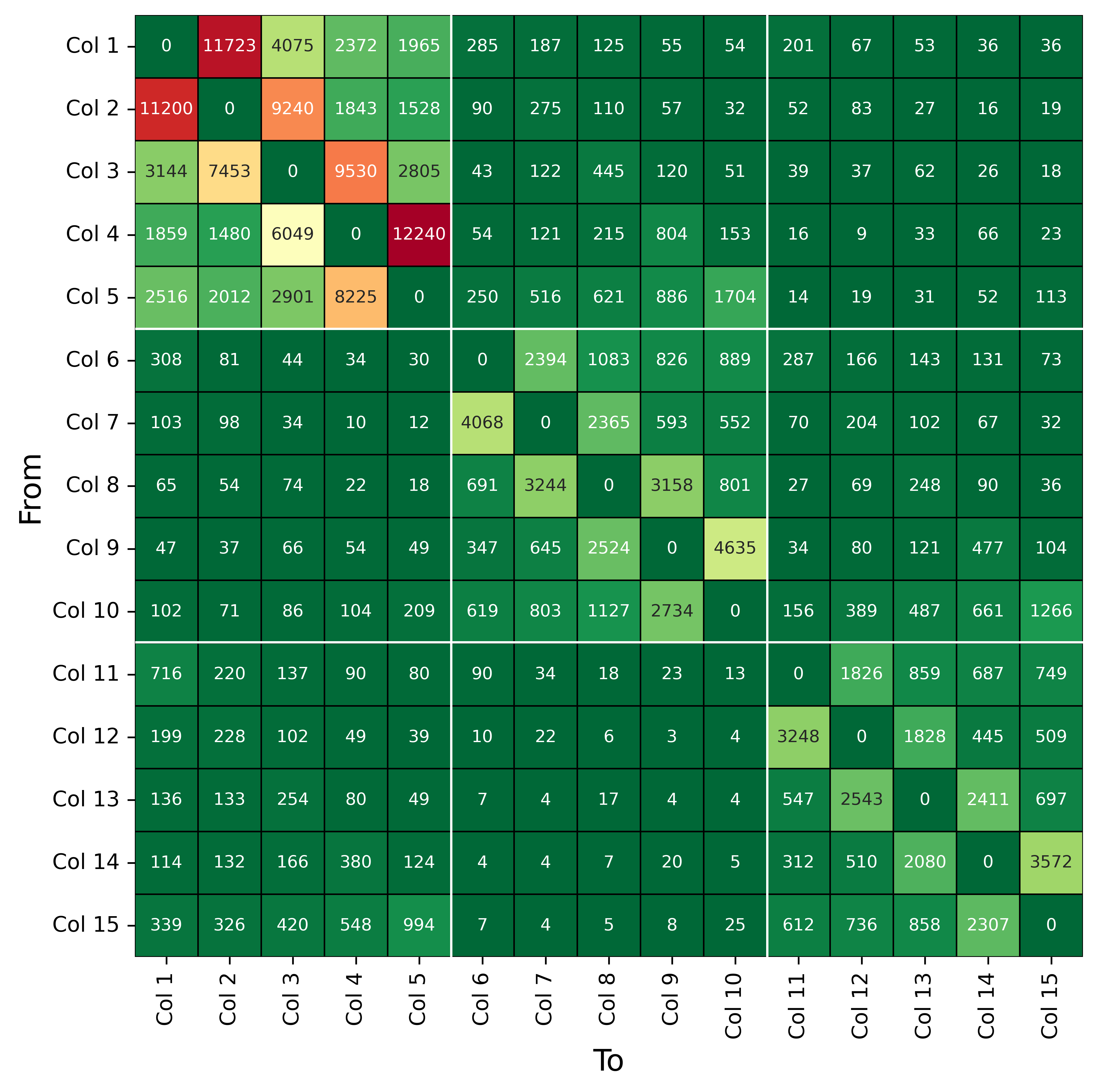} 

        \includegraphics[clip, trim =0mm 180mm 17mm 0mm, width=0.697\linewidth]{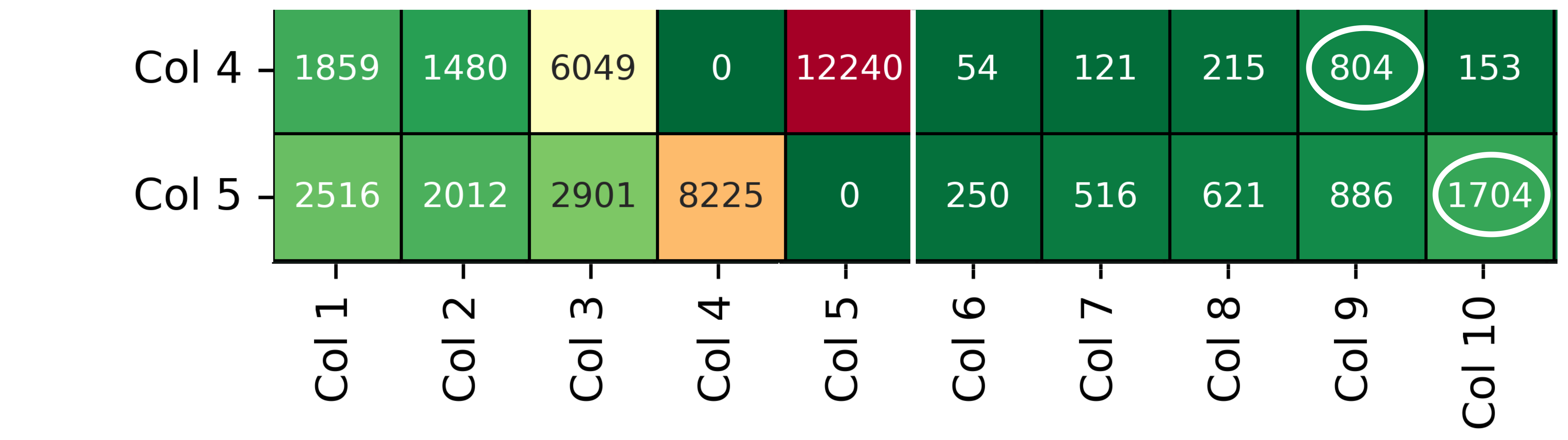} 
        \caption{i) users swipe right rather than left and from the rightmost movie, ii) after swiping, users examine the same position (col+5), and iii) re-examining items is more likely than swiping.
        \vspace{.1cm}
        }

    \end{subfigure}
    \begin{subfigure}{0.20\textwidth}
    \centering
        \includegraphics[clip, trim =75mm 0mm 0mm 0mm, height = 3cm]{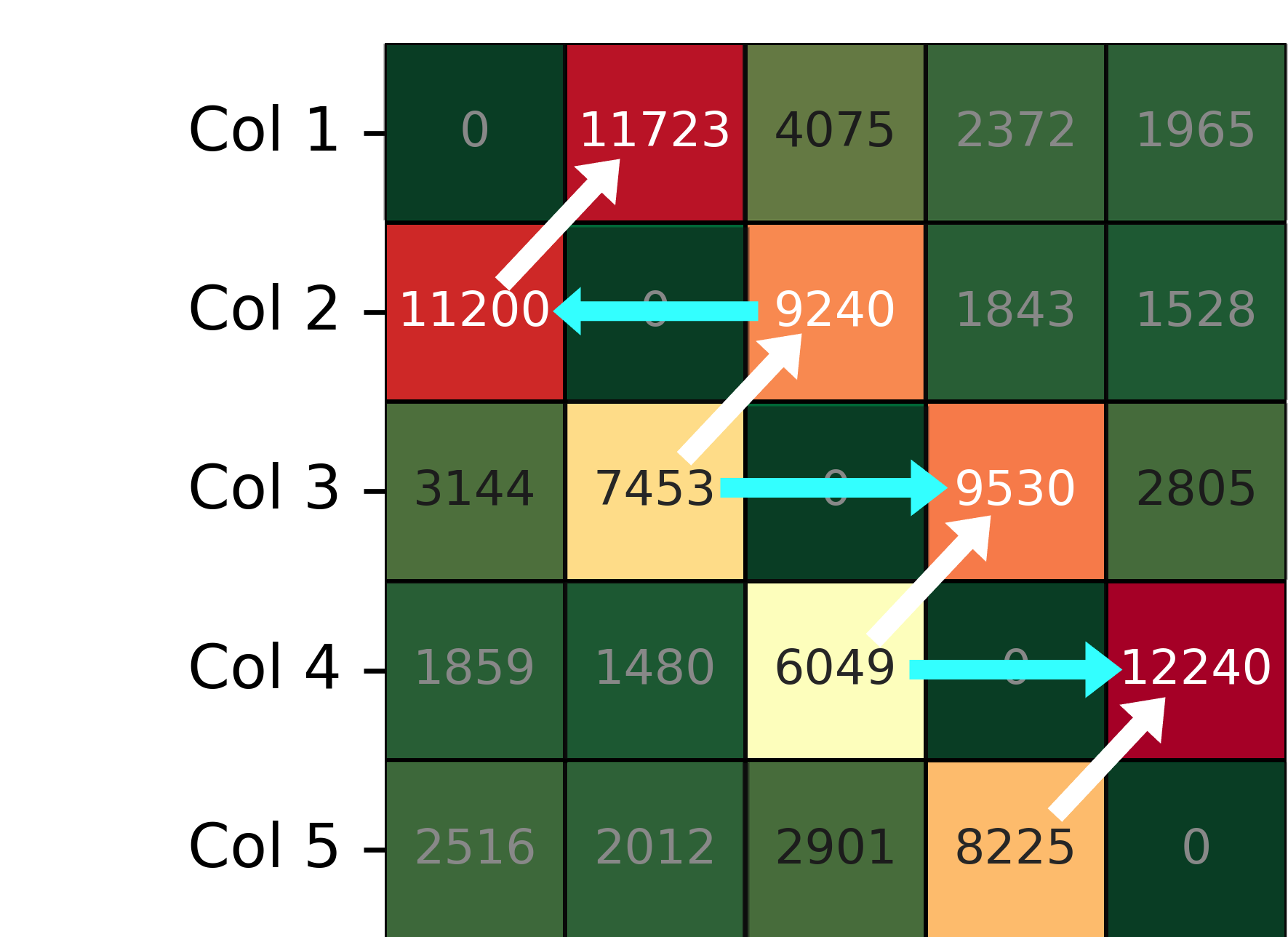}
        \caption{On 1st set of movies, users tend to browse left-right except for Movie 2. The upper diagonal is left-right browsing and the lower diagonal is right-left browsing.   
        \vspace{.1cm}
        }
        \includegraphics[clip, trim =75mm 0mm 0mm 0mm, height = 3cm]{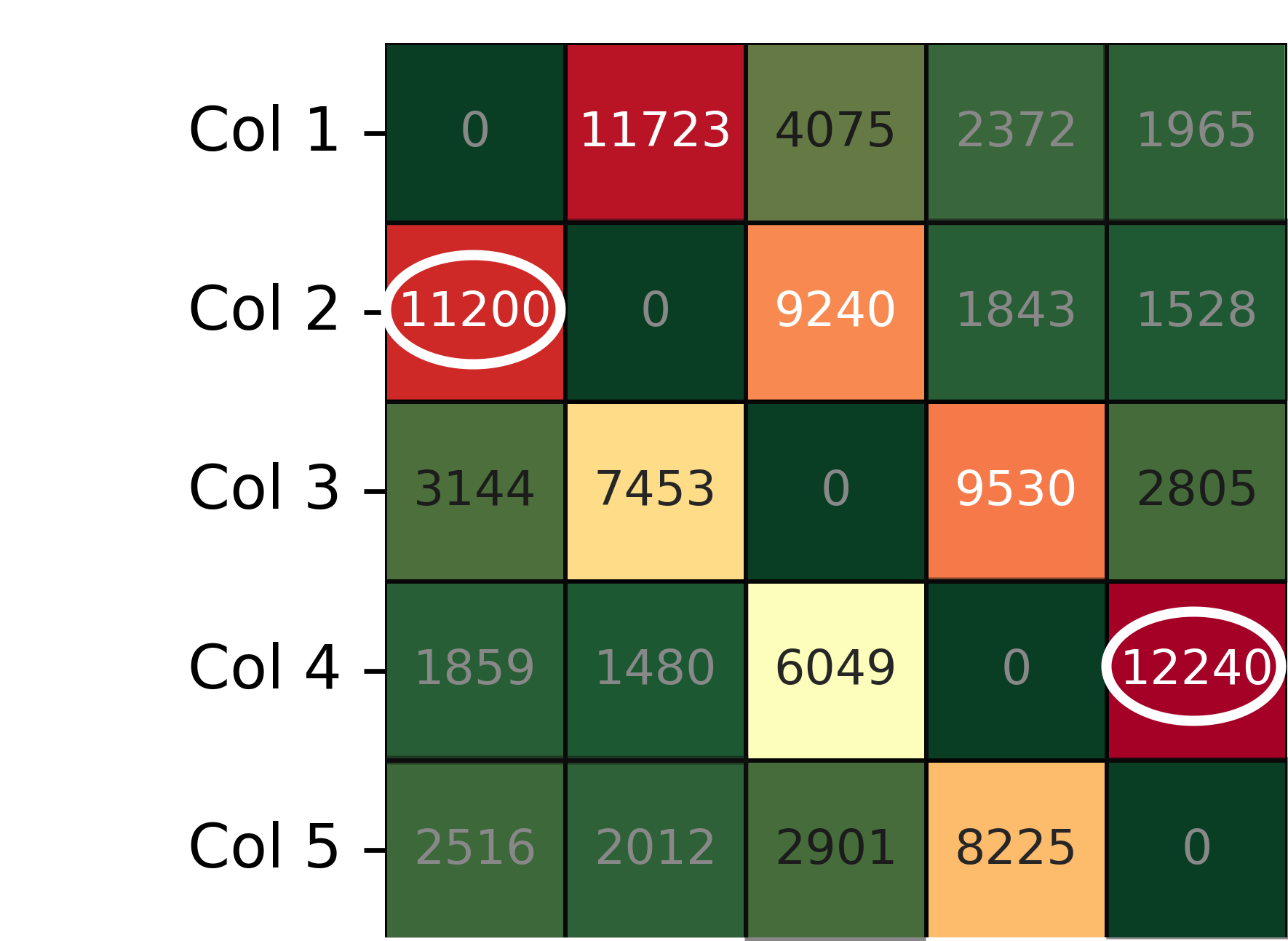}
        
        \caption{High transitions to edge. 
        \vspace{.1cm}
        }
        \includegraphics[clip, trim =170mm 0mm 0mm 0mm, height = 3cm]{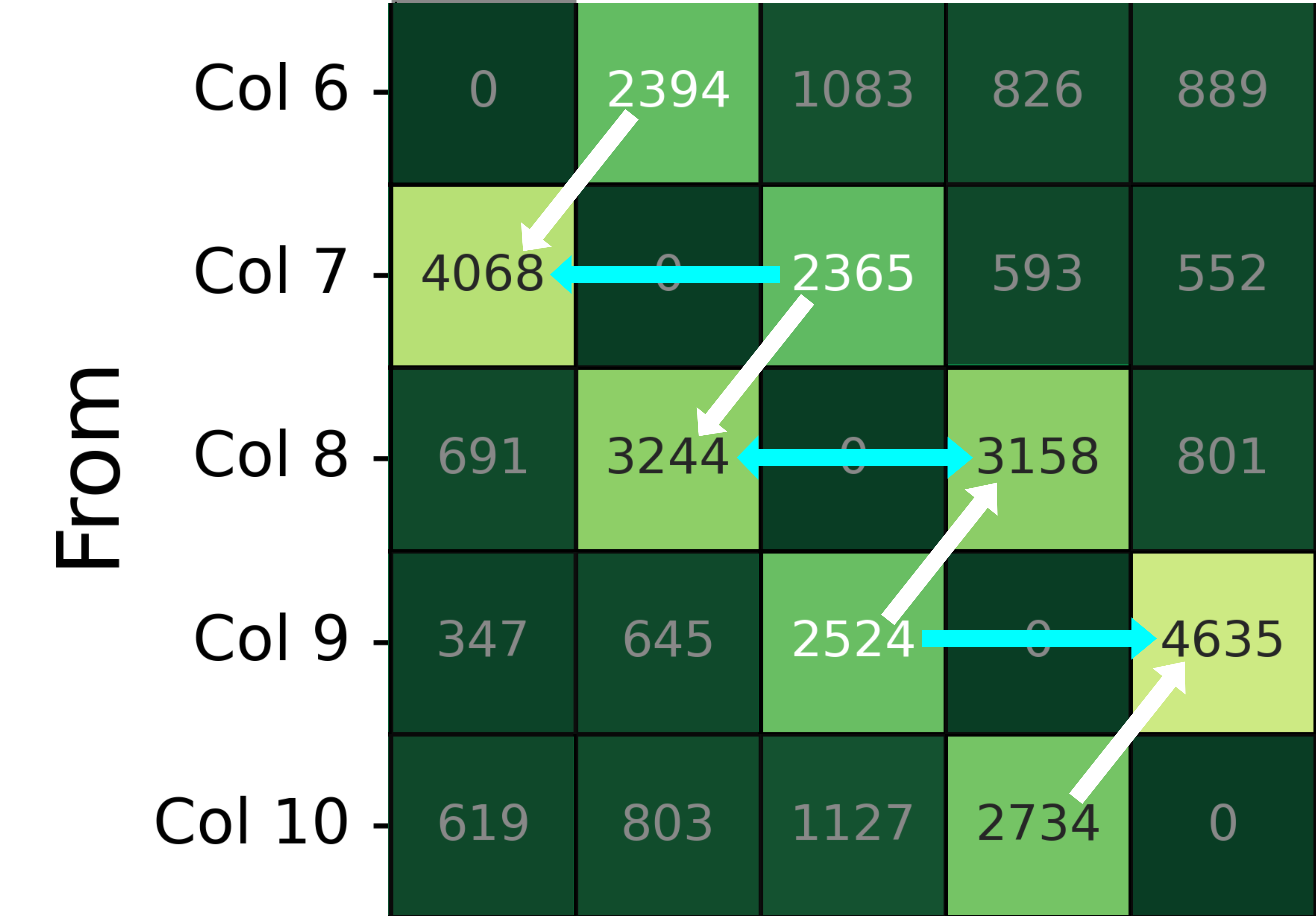}
        \caption{After swiping, users browse from center to edge.
        \vspace{1.3cm}
        }
    \end{subfigure}
    
    \caption{Column fixation transitions matrix for all users and free-browsing screens. It shows the number of instances users examined an item column (one of the 15 movie rank positions independent of genre) and afterwards examined a different item. All fixations on the background, genre label, and swipe buttons are ignored. Note that Columns 1 through 5 are the first set of movies, Columns 6 through 10 are the second set of swiped movies, and Columns 11 through 15 are the third set. When swiping, the columns shift 5 places (e.g. for a right swipe, Column 1 is replaced by Column 6). Thus, the top left, center, and bottom right areas of the matrix independently visualize a user transitioning without swiping. (a)-(d) visualize the main results in subareas of the matrix.}
    \label{fig:col_transitions}
\end{figure*}

\subsection{RQ 3: How do users transition across items in a carousel (column-to-column)?}

Similar to the row transitions, \cref{fig:col_transitions} shows the number of fixation transitions from one item column (position 1-15 of a movie independent of the carousel genre row) to another item column. The matrix computation is similar to the row transition matrix, but only fixations on an item (movie poster) are used while all others are ignored. Therefore, the matrix shows how users transition left or right from movie to movie including the 2nd (Columns 6 through 10) and 3rd (Columns 11 through 15) set of movies that must be swiped to.

We begin by examining item transitions on the initial/unswiped set of movies (upper left corner of the matrix).  \cref{fig:col_transitions}b generally shows a left-right browsing behavior with arrows pointing to the top diagonal of left-right transitions, which is in line with a F-pattern left-right reading behavior on the initial items. 

This left-right browsing behavior is broken when a user is at Movie 2, where they are more likely to move left back to Movie 1 than they are to move right to Movie 3.  Similarly,  \cref{fig:col_transitions}c shows a pattern (that repeats in the middle and bottom right subareas) of large transitions values to the edge movies. We hypothesize that these results are due to users alternating between left-right and right-left browsing behaviors after initial examination for re-examining the items. In the case of the initial/unswiped items users are very likely to browse left-right and then will browse right-left on re-examining with a high likelihood to go to the edge, which not only finishes the left-right or right-left behavior, but also allows for swiping.

Extending the analysis to swiped set movies (center and bottom right areas of the matrix) the pattern of high transitions to the edge in \cref{fig:col_transitions}c is repeated. Similarly to the initial set of of movies, users are likely to re-examine the swiped movies. Moreover, \cref{fig:col_transitions}d shows that after swiping users no longer show a clear dominant general browsing behavior of left-right or right-left, but instead a general browsing behavior towards the edge.  %

To better understand this edge directed browsing behavior in swiped movie sets, we inspect how a user swipes and what item is examined first after swiping. \cref{fig:col_transitions}a provides an example of transitions after swiping. Users are much more likely to swipe right rather than left. Moreover \cref{fig:col_transitions}a shows that users are more likely to swipe from the last movie position (Col 5) than the second to last (Col 4). This pattern continues with very little transitions from the center or left movies to next swiped movies.  After the swipe,  users tend to fixate on the movie that slides into the position where they were previously fixating (see \cref{fig:col_transitions}a). %
This behavior of swiping on the 5th position and fixating on the new item in the same 5th position explain the reverse right-left browsing behavior observed in the preliminary results of the dataset paper~\cite{deleon_et_al_2025}. It also motivates one of our suggested positions for ranked items in carousel interfaces (see Section \ref{takeaways}). To explain the edge directed browsing behavior, the initial examination of the swiped set of items is likely done right-left, but after initial examination there is no prominent left-right or right-left behavior. As users re-examine the items, they are more likely to move the edges to swipe left to return to a previous interesting item or swipe right to continue exploring.%

In terms of non-adjacent or skip transitions within the same set of movies, these are more common than row skipping (when compared to their respective adjacent transitions). However, this is not surprising as glancing between the movies already presented is not as \textit{costly} as scrolling. Moreover, users are most likely to skip from the edge movies and are more likely to perform a skip transition in swiped sets of movies. This may be a possible explanation for the center to edge browsing behavior observed in swiped item sets. Finally, the top left, central, and bottom right areas of the matrix contain the majority of transitions meaning that a user is much more likely to re-examine movies already present than swipe to new movies. 

In conclusion, the column transitions reveal that:
\begin{enumerate*}[label=(\roman*)]
\item On the initial, unswiped set of items, users initially browse left-right and also generally move left-right when re-examining items,
\item users on the near-edge items are likely to transition to the edge (if swiped or unswiped),
\item users primarily swipe using the right swipe button after observing the last right-hand item,
\item after swiping users examine the same item position as before,
\item on swiped sets of items, users initially browse right-left and then generally browse towards the edges when re-examining items,
\item users are most likely to skip from edge movies and skipping is more likely in swiped item sets, and
\item users are more likely to re-examine the 5 visible movies than swiping to see the next set. 
\end{enumerate*}

\heading{Joint row-column transitions}

\begin{figure}[!tb]
    \centering
    \includegraphics[width=0.7\linewidth]{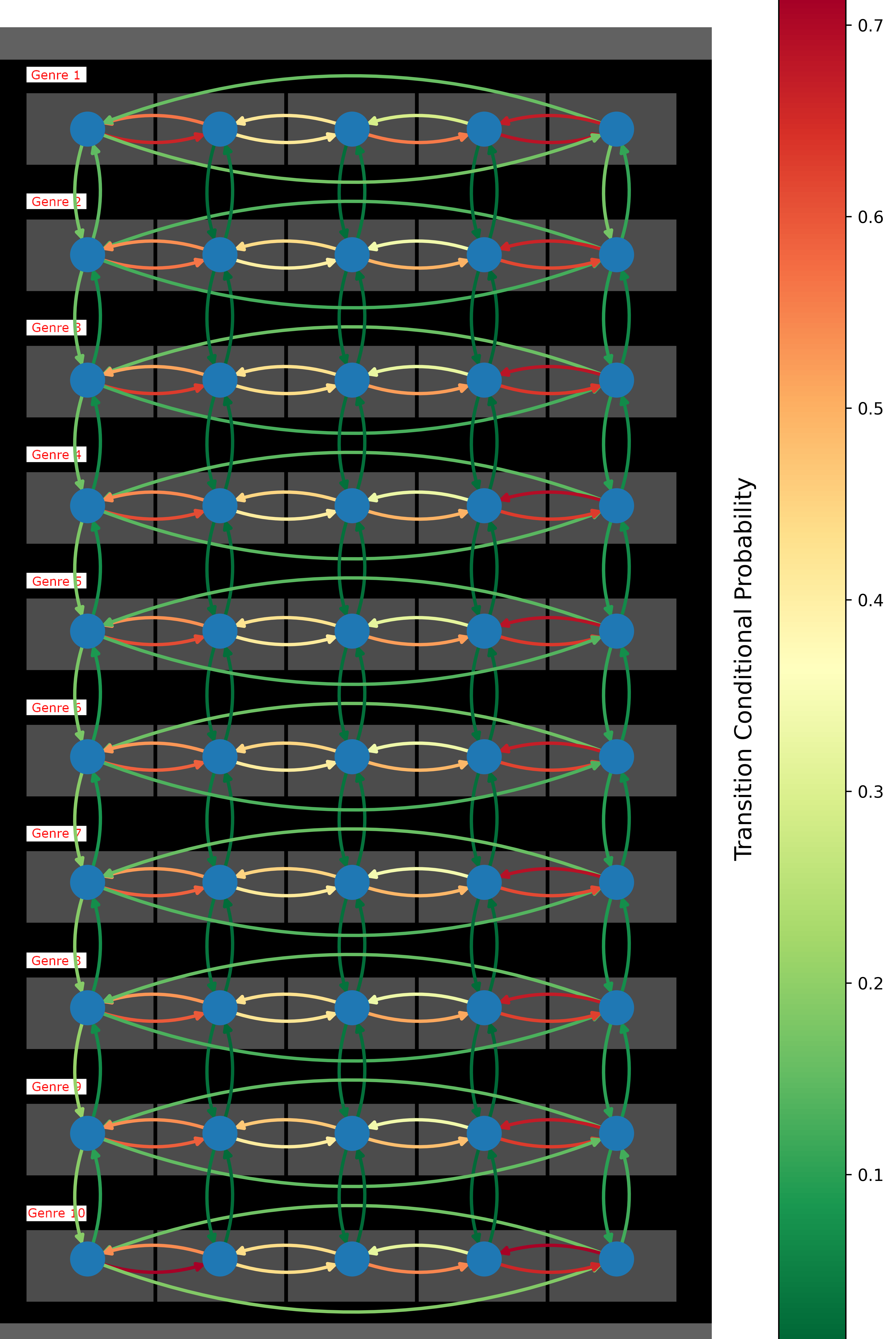} 
    \caption{Joint row-column fixation transitions for all users and free-browsing screens. Each arrow is a item-item transition through eye movement, colored  by the transition conditional probability (given the starting item position). Transitions are aggregated across swiping to be in line with the true browsing path of users. This means that in Genre 1 the arrow from the first item to the second item represents the transitions in Row 1 for item 1-2, 6-7, and 11-12, while the arrow from the first item of Genre 1 to the first item of Genre 2 represents all of the possible transitions from Row 1 item 1, 6, or 11 to Row 2 item 1, 6, or 11.}
    \label{fig:grid_transitions}
\end{figure}

To better visualize the results of RQ 2 and 3, we combine both row and column transitions. \cref{fig:grid_transitions} shows the joint row-column transition matrix on a to-scale copy of the interface. All initial and swiped items are aggregated, since a swipe fully replaces the 5 items with a new set of 5 items. We also include transitions from the first to the last item (long horizontal arrows) as a sanity check. This confirms that users are very unlikely to go back to the first, leftmost item after swiping (repeating the left-right browsing behavior) and that users are initially browsing right-left on swiped sets of items. 

Additionally, the figure supports the previous results for row and column transitions. For row transitions, a general top-down browsing behavior is seen with a higher probability to transition down. A new insight can be seen that these vertical transitions are likely to happen from edge movies and very unlikely from central movies. Moreover, the top 2 row bias can also be seen as the most prominent bottom-up transitions are row 2-1.

\begin{figure*}[!tb]
\centering
    \begin{subfigure}{0.55\linewidth}
        \includegraphics[clip, trim =65mm 65mm 0mm 0mm, width=1.0\linewidth]{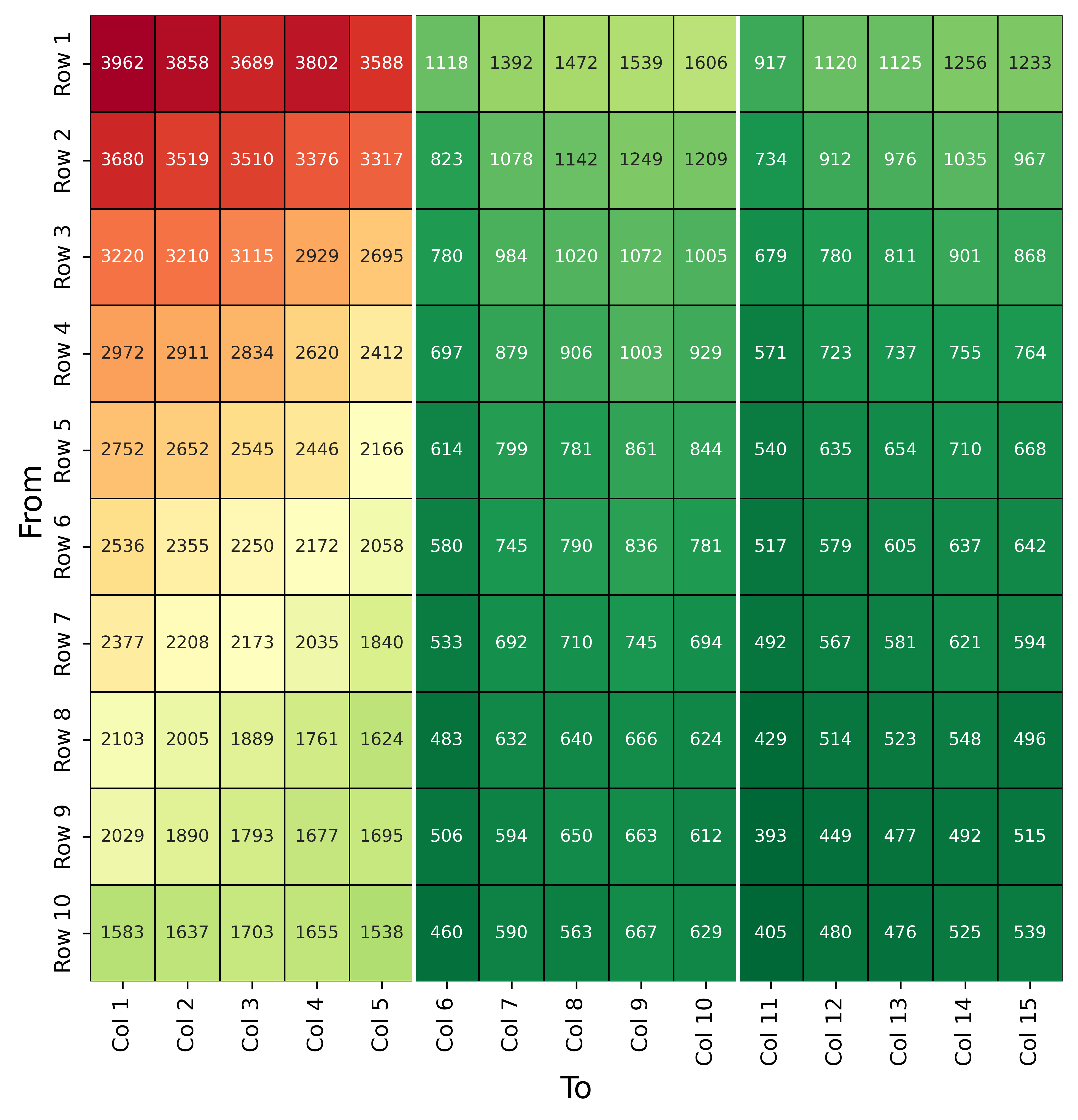}

    \end{subfigure}
    \begin{subfigure}{0.44\linewidth}
    \centering
        \includegraphics[height = 3.5cm]{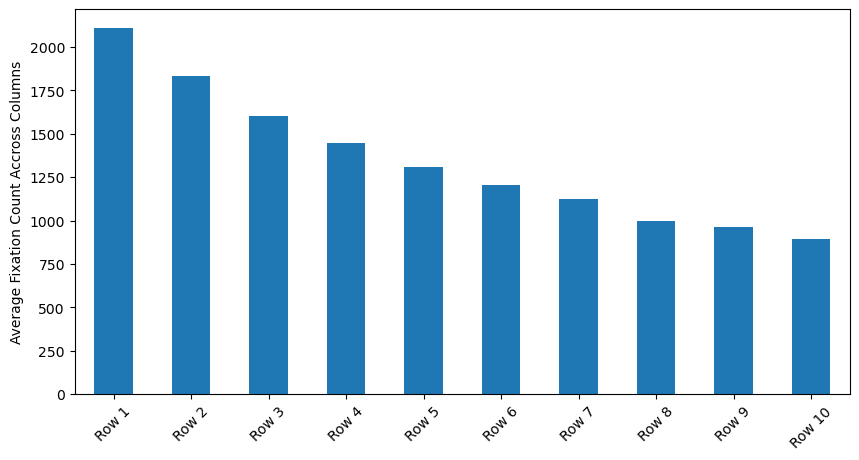}
        \caption{Average fixation count on rows. To summarize how users fixate on an entire row, we average across all the columns in the heat map. 
        \vspace{.3cm}
        }
        \includegraphics[height = 3.5cm]{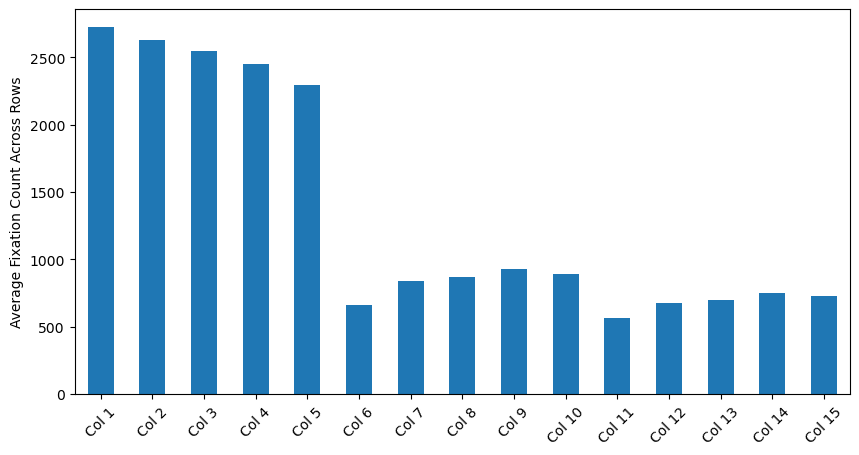}
        
        \caption{Average fixation count on columns. To summarize how users fixate on an item position regardless of row, we average across all the columns in the heat map. 
        }
        \vspace{.3cm}
    \end{subfigure}
    
        \caption{Fixation heat map of each movie item for all users and free-browsing screens. We remove all repeat fixations on the same movie and count only fixations transitions to other movies and the initial movie fixation. Only fixations for Movie AOIs are used.}
    \label{fig:heat_map}
\end{figure*}

For column or item-item transitions, we see a general browsing behavior of center to edge, as swiped and unswiped item sets are mixed together. This is in line with the fact that users are generally traversing these items left-right and right-left when re-examining items (more prominent than the initial browsing/examination of the items). Of note, there are high probabilities of transitioning to near edge movies, which is inline with general center to edge browsing behavior and that users are more likely to re-examine items than swipe. In the next section, we show that there is a surprising bias for a near edge movies in swiped sets movies.

\heading{Empirical Fixation Propensities} In \cref{fig:heat_map} we recreate the fixation heat map from the dataset paper \cite{deleon_et_al_2025} showing the fixation counts for each item and only counting the initial item fixation and fixations to new items (i.e. consecutive fixations on the same item are removed). This provides a heat map that is more representative of how users transition and less dependent on the individual item biases that can encourage a user to fixate on it multiple times in a row, especially when reading the description of the movie. 

The heat map color gradient supports left-right browsing behavior  on the unswiped  set of movies and right-left browsing behavior on the swiped set of movies. However, surprisingly, the column averages (across all rows) in \cref{fig:heat_map}b show that column 9 (movie in the 4th position or second last after one swipe) has more fixations than column 10 and similarly with the last set of movies (column 14 > column 15). \cref{fig:heat_map}a provides further evidence for top-down browsing and the top 2 row bias, due to the exponential curve of the bar graph (i.e., Row 1 and Row 2 receive a higher proportion of fixations). These results are the motivation for our proposed ranking orders below in \cref{takeaways}.

\subsection{RQ 4: How does genre preference impact row transitions?}
User genre preference information, gathered by pre-surveys, is included in the dataset. Each user was asked to provide their preference in 3 different methods for the 10 different genres: marking preferred / non-preferred for each, 1-star to 5-star rating for each, and pick the top or most preferred genre. For each of the preference methods, we analyze their impact on row transitions (from one genre carousel to another). 

\begin{figure}[tb!]
    \centering
    \begin{subfigure}{0.25\textwidth}
        \includegraphics[width=\linewidth]{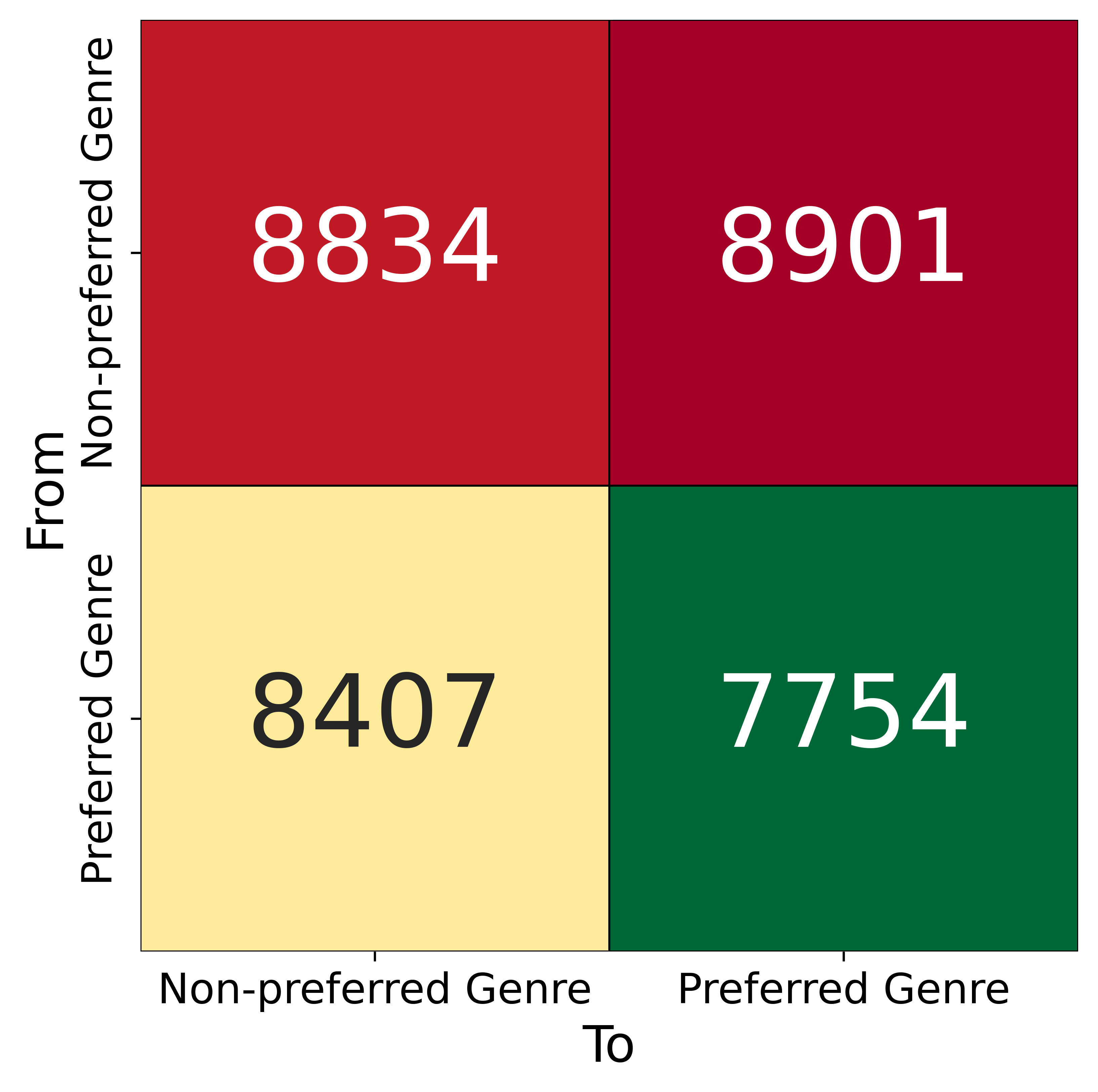} 

        \label{fig:prefgenre_transitions}
    \end{subfigure}
    \begin{subfigure}{0.25\textwidth}
    \centering
        \includegraphics[width=\linewidth]{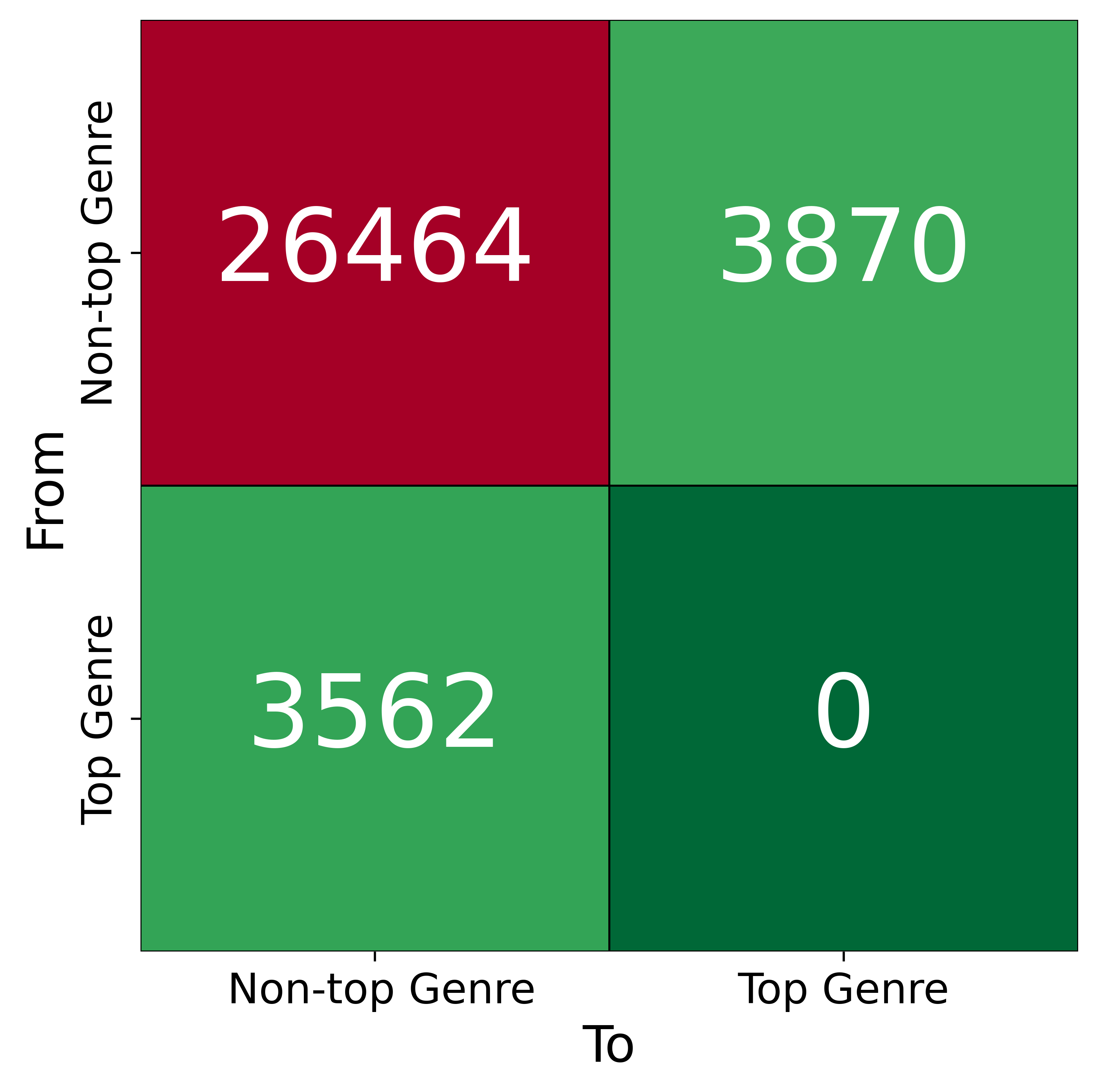}
    \end{subfigure}
    \caption{Genre preference row fixation transitions matrices for all users and free-browsing screens. Similar to \cref{fig:row_transitions}, but rows are labeled by the users' recorded preference as either preferred/non-preferred or top/rest. The average user selected 4.7 preferred genres (5.3 non-preferred genres).}
    \label{fig:preferred_transitions_together} 
\end{figure}

\heading{Preferred genres vs. non-preferred genres}
\cref{fig:preferred_transitions_together} shows the preferred  vs. non-preferred genre row transitions for fixations. Given either row, the transition to the same type of row is slightly less than to the other type (preferred to preferred < preferred to non-preferred as well as non-preferred to non-preferred < non-preferred to preferred). This result is most likely due to the balanced ratio of on average 4.7 preferred genres to 5.3 non-preferred genres (i.e. once given a row, one of that type has been removed as a possible transition target). A clearer result is present in the reverse transitions, with preferred to non-preferred having 7,754 counts vs. non-preferred to preferred having 8,901 counts. This shows that users are more likely to transition to preferred genres despite there being a higher average prevalence of non-preferred genres (5.3 vs. 4.7) to preferred genres. As a follow-up, a Wilcoxon singed-rank test indicated that users made significantly more transitions from non-preferred to preferred than preferred to non-preferred ($w=3205.500, p<0.001$).

\heading{Top genre vs. rest}
\cref{fig:preferred_transitions_together} presents the fixation transition matrix for the top genre vs the rest. There is only one top genre, so there are no possible transitions from a top genre to another top genre. Examining reverse transitions, we see that transitions to the top genre are more likely than transition out of the top genre. This shows that users were more likely to transition to the top genre and also less likely to leave it. A Wilcoxon singed-rank test indicated that users made significantly more transitions from non-top to top than top to non-top ($w=2897.500, p<0.001$).

\begin{figure}[tb!]
    \centering
    \includegraphics[width=0.5\columnwidth]{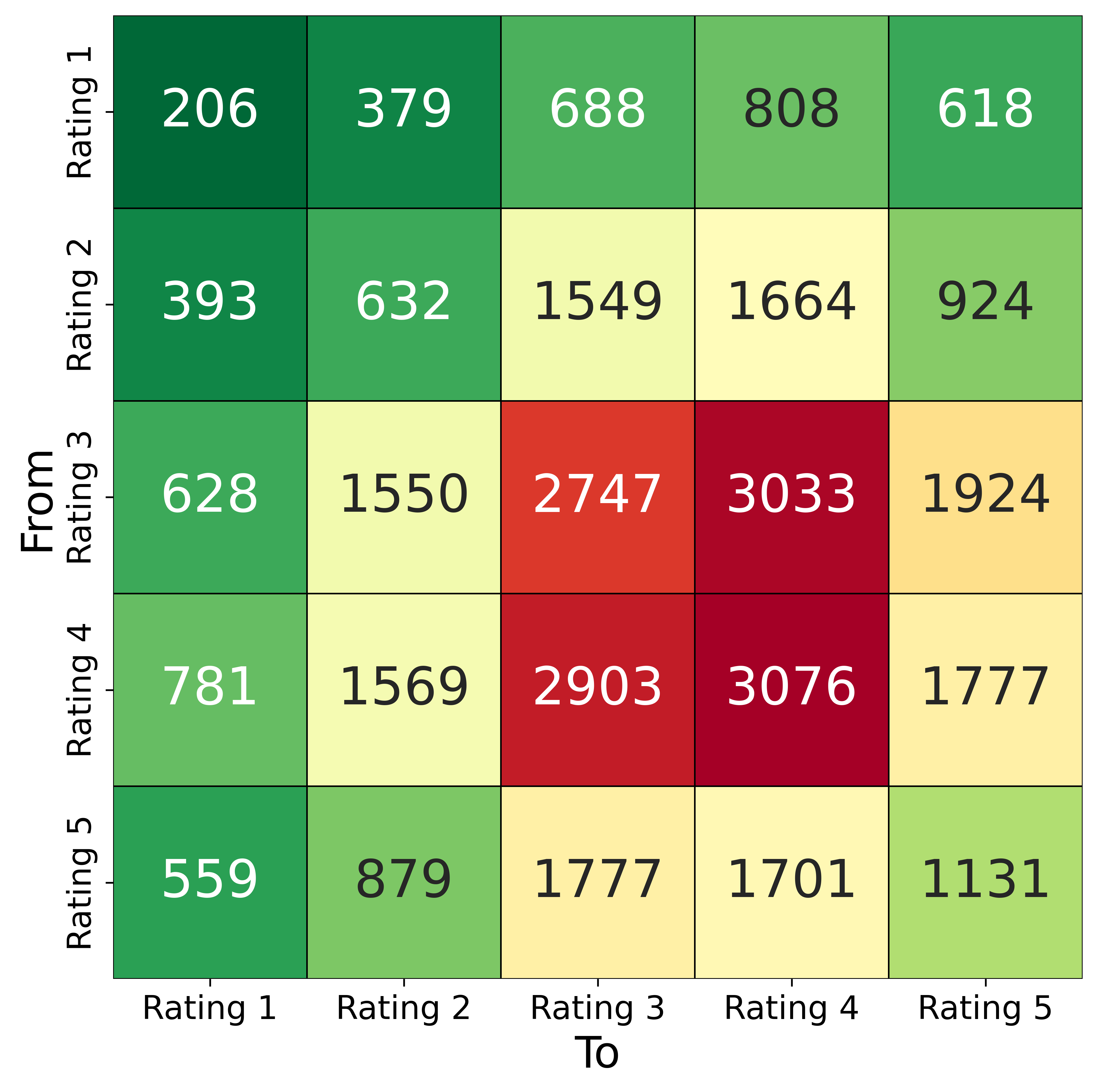} 
    \caption{Genre rating row fixation transitions for all users and free-browsing screens. The ratios (across all users) of ratings 1 to 5 respectively were 1:1.8:3.4:3.4:2.1.}
    \label{fig:rating_transitions}
\end{figure}

\heading{Genre rating transitions}
Finally in \cref{fig:rating_transitions}, the genre rating row transitions for fixations are shown. In general, users are more likely to transition to/from ratings 3 or above than ratings 1 and 2. This is observed by the higher amount of transitions tending to the right of the matrix. However, transitions to rating 5 carousels are not the most prominent due to lower prevalence of such highly rated genres. A Wilcoxcon singed-rank test indicated that users made significantly more transitions from lower ratings to higher ratings than higher ratings to lower ratings ($w=3246.000, p<0.001$).

The above analyses of preference suggest that genre preference impacts carousel browsing behavior (despite the random ordering of genres across all screens). All 3 methods of measuring genre preference were found to be statistically significant in favoring preferred, top, and higher rating genres. While these results are not surprising, they point to the importance of topic preference when considering how users browse carousels, user modeling, and design of carousels.

\section{Takeaways for carousel recommenders}
\label{takeaways}

Based on the analysis and insights learned in answering the above research questions, we provide a series of suggestions and practical takeaways for improving recommender systems for carousel homepages (free-browsing assumption). These takeaways are empirically supported by the analyses of this dataset, though their validation and impact on deployed recommender systems remains future work. 

\begin{figure}[tb!]
    \centering
    \includegraphics[width=\columnwidth]{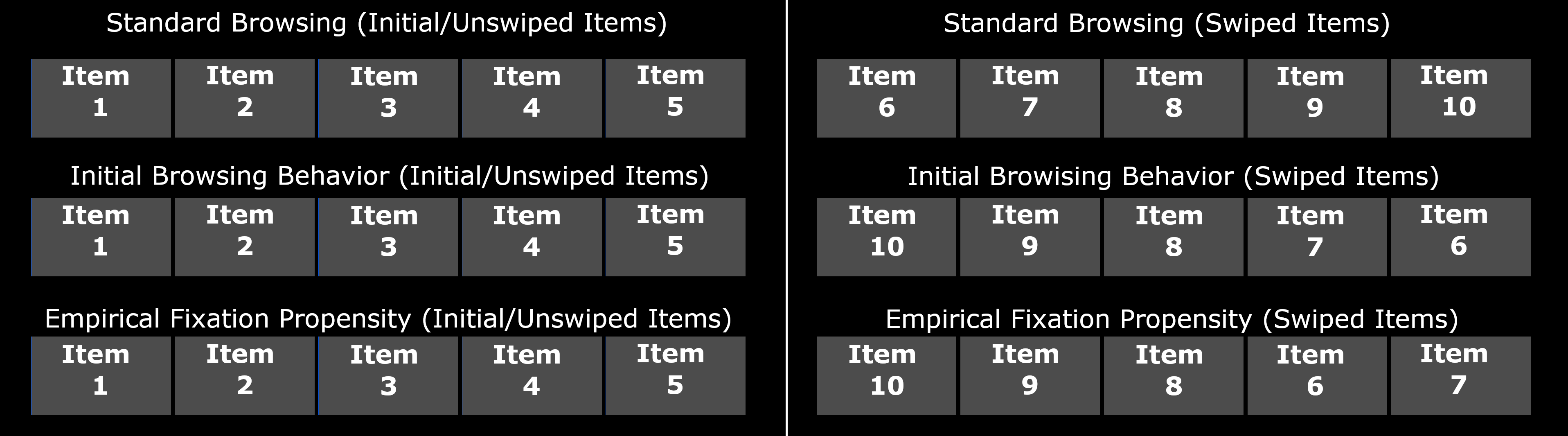} 
    \caption{Proposed ranked item positioning}
    \label{fig:ranking_order}
\end{figure}

\heading{Ranked item positioning}
The most surprising and relevant result of the eye tracking analyses is how users browse after swiping a carousel. In \cref{fig:ranking_order} we show how ranked items are positioned traditionally or what we call standard browsing (only left-right browsing), which is in line with the analyses for the initial/unswiped items. However, our results show that after a swipe, users initially browse right-left starting from the same item  that they previously were examining before the swipe. Thus, in \cref{fig:ranking_order} we propose that after a swiped sets of items should be ordered from right to left. Another possible ordering can be taken from the empirical fixation results of \cref{fig:heat_map}, what we call empirical fixation propensity. This ordering is similar keeping the right-left browsing on swiped items, but prioritizes the 4th position as the highest ranked item of swiped sets. %

\heading{Top 2 row bias} Most carousel recommender systems are likely to be designed with most relevant carousel rows (or rows that need to be seen for business constraints) at a top row position rather than a lower row position. Designers should pay special attention to the top 2 rows when considering relevance or business constraints. For click modeling, the observed back and forth transitions between row 1 and row 2 violate the top-down browsing assumptions of many click models. A possible solution is to design more complex click models that are able to account for this type of browsing behavior. %

\heading{Row skipping is generally rare} Users tend to skip rows rarely (1:10 ratio of skip to non-skip) except for the first and last row where skipping is more likely. This provides insights on how to build user/click models under the free-browsing setting. The only carousel click model proposed does not fit this behavior as it models a user finding only one relevant topic and only browsing the items in that topic \cite{rahdari_ranked_2022}. Despite not being matched to the actual user behavior, it is still possibly correct in predicting the empirical click positions. This is something we will explore in a follow up click modeling work. Moreover, an individual or clustered analysis of user browsing behavior could provide deeper insights into skipping that are not present in the aggregate analysis (e.g. do some users tend to skip while others do not?).

\heading{Topic preference impacts browsing} The 30 screens of the user study were designed so that each genre was randomly placed, but at the same time equally present in the 3 initial presentation locations and equally present in each of the 10 possible row positions. Despite this randomness (of row positional bias), results still showed higher transitions to carousel rows that were evaluated positively through preference metrics. For this reason, it can be very beneficial to learn the users' preferences for topics/genres rather than primarily focusing on item preference. Future work can be done examining how preference impacts clicks and the time spent or number of fixations on a row. 

\heading{Positional bias} The above results and takeaways show the importance varying positions (row or column) have on how users browse carousel and how likely they are to observe a certain item. For this reason, accounting for positional bias in carousel interfaces (e.g. debiasing clicks based on position) may be much more necessary than other types of interfaces. The multiple empirical figures presented herein can be used as estimations for initial bias values. 

\section{Limitations}
\label{Limitations}

Since we base our analyses on the user study and the dataset published in~\cite{deleon_et_al_2025}, some of the limitations mentioned therein also apply to our work. Additionally, there are possible limitations of our analyses as mentioned below.

\heading{User Study limitations} The above average socioeconomic and economic level of the participants of the user study may not be representative of the average movie streaming service user. This is likely to impact preferences, but may or may not impact general browsing behavior. Moreover, this study did not consider visual bias of the posters that may impact browsing through color and content. This is addressed in our analyses by aggregation (over screens/partici\-pants), so that individual screen and user differences are less impactful.  

\heading{Dataset limitations} Small inaccuracies present in eye tracking data may have led to incorrect labeling of an AOI for some fixations. While the size of the dataset is very large for an eye tracking study (87 users and 30 screens), further research needs to be done to be more representative of the large amount of users globally using carousel interfaces, particularly in an industrial setting. 

\heading{Analysis limitations} The results of our analyses may be somewhat limited to the study interface design (although it opted for a simple, familiar interface close to Netflix) and the movie domain. It is possible that different domains with different purposes or topics may have different browsing behaviors. For example, the browsing behavior of a user on an e-commerce platform with carousels looking for home goods may be different as the context and task are not entertainment related. More eye tracking user studies are needed for domains outside of movie recommendation. Additionally, our results on swipe behavior are limited to the common swiping behavior on desktop to replace all items with a new set on swipe. There are some interfaces that on swiping only slide one item off the page and show one additional, new item. While such swiping behavior should be also be studied with eye tracking, we hypothesize that the browsing behavior would be more intuitive. Finally, in terms of generalizability of our results to similar movie recommender interfaces, but with slight differences in presentation (e.g. differing amounts of items or number of carousel lists), our results and the general patterns are likely to hold as these are minor differences that should not affect the browsing strategies or behaviors of the user. 

\section{Conclusion}
Despite the prevalence of carousel, there is a gap in the literature of how users actually browse these complex interfaces. In this work, we provide the first extensive eye tracking analysis of browsing behavior in carousels to help answer this open question. We began by examining how users start browsing in the top left area of the initial rows. Followed by a row-to-row, column-to-column, joint row-column transition, and genre preference analysis with three main results:
\begin{enumerate*}[label=(\roman*)]
\item after swiping users browse right-left starting from the previously observed item,
\item there is a top 2 row bias, and
\item topic preference does impact browsing despite random row ordering.
\end{enumerate*}
Finally, we provided practical suggestions for carousel recommender designers and proposed an improved ranked item positions for carousel interfaces. Also we hope to encourage other researchers to continue work on exploring the complexity of carousel browsing behaviors.

As mentioned above, one of our future works will be to develop a carousel click model based on the eye tracked browsing behavior as well as the general insights learned on how users browse. This will be only one of the many ways that a carousel user model can be designed based on the insights herein, but will allow us to evaluate how closely it fits to the real-world click data. These user models can also be used to further analyze and classify browsing behavior of differing user groups along with a joint analysis of user perception, satisfaction, and decision-making measures that are also available in the dataset.  Extending beyond browsing behavior, the eye tracking data can be further analyzed using fixation durations or other more fine-grained gaze metrics for building richer recommenders that are either gaze-based (treating gaze as feedback) or gaze-informed. Not only do we hope to encourage research in carousel-based interfaces, but more broadly on the potential of gaze signals in recommender systems.%
\section{GenAI Usage Disclosure}
Github Copilot was used as a coding assistant with human review and validation of generated code. GenAI was not used in writing of the manuscript.

\begin{acks}
This work was supported by Eyes4ICU, a project funded by the European Union under the Horizon Europe Marie Sk\l{}odowska-Curie Actions, GA No. \href{https://doi.org/10.3030/101072410}{101072410} and by LorAI - Low Resource Artificial Intelligence, a project funded by the European Union, GA No. \href{https://doi.org/10.3030/101136646}{101136646}. We would like to acknowledge the support of Eye Square and Blickshift Analytics GMBH \cite{Blickshift_2024} for its support with software and eye tracking analysis. 

\end{acks}

\bibliographystyle{ACM-Reference-Format}
\bibliography{references.bib}
\end{document}